\documentclass[twocolumn,superscriptaddress]{revtex4-1}
\usepackage[dvips]{graphicx}
\usepackage{type1cm}
\usepackage[usenames,svgnames]{xcolor}
\usepackage{color}
\usepackage{url}
\usepackage{natbib}
\usepackage[colorlinks=true,linkcolor=blue,citecolor=blue,breaklinks=true]{hyperref}
\usepackage{amsmath,amssymb,bm,amsthm}
\usepackage{newtxtext,newtxmath}
\usepackage{braket}

\begin{document}
\nocite{apsrev41control}

\title{Metastability associated with many-body explosion of eigenmode expansion coefficients}

\author{Takashi Mori}
\affiliation{
RIKEN Center for Emergent Matter Science (CEMS), Wako 351-0198, Japan
}

\begin{abstract}
Metastable states in stochastic systems are often characterized by the presence of small eigenvalues in the generator of the stochastic dynamics.
We here show that metastability in many-body systems is not necessarily associated with small eigenvalues.
Instead, many-body explosion of eigenmode expansion coefficients characterizes slow relaxation, which is demonstrated for two models, interacting particles in a double-well potential and the Fredrickson-Andersen model, the latter of which is a prototypical example of kinetically constrained models studied in glass and jamming transitions.
Our results provide insights into slow relaxation and metastability in many-body stochastic systems.
\end{abstract}
\maketitle

\section{Introduction}
Metastability in stochastic systems is ubiquitous in nature.
In particular, it is often associated with first-order phase transitions~\cite{Langer1969, Langer1974, Penrose1995}, (spin) glasses~\cite{Bray1980, Biroli2001, Berthier2011}, protein folding~\cite{Baldwin2011}, and so forth.
Despite its importance, the definition of metastability is quite subtle.
In mean-field models, there is a static characterization of metastable states, i.e., local minima of the free energy as a function of order parameters~\cite{Griffiths1966}.
For non mean-field models, however, a metastable state has a finite lifetime even in the thermodynamic limit~\cite{Lanford1969}, and hence we should treat metastability as a timescale-dependent dynamical concept~\footnote{See Refs.~\cite{Langer1969, Penrose1995} for attempts to give a static characterization of finite-lifetime metastable states associated with first-order phase transitions via an analytic continuation of the thermodynamic function.}.

Since the generator $G$ of the master equation contains full information on the stochastic dynamics, it is natural to expect that metastability is dynamically characterized by some special properties of eigenvalues and/or eigenvectors of $G$.
Indeed, the real part of an eigenvalue of $G$ gives the decay rate of the corresponding eigenmode, which suggests that metastability is characterized by the presence of small eigenvalues separated from the other large eigenvalues~\cite{Gaveau1998, Bovier2000, Biroli2001} (see Refs.~\cite{Macieszczak2016, Rose2016, Rose2020} for quantum systems).
Such a spectral characterization of metastable states is appealing since it is purely dynamical; we do not have to rely on thermodynamic notions such as the (free) energy landscape, which is difficult to define precisely except for mean-field models~\cite{Frenkel2013}.

In this paper, we revisit this problem.
We show that metastability in \emph{many-body} stochastic systems does not necessarily accompany small eigenvalues.
By analyzing two simple models, i.e. interacting particles in a double-well potential and the Fredrickson-Andersen (FA) model~\cite{Fredrickson1984}, we demonstrate that slow relaxation in those models is characterized by \emph{many-body explosion} of eigenmode expansion coefficients, which was recently studied in the context of open quantum systems~\cite{Mori2020_resolving, Haga2021} (see also Ref.~\cite{Bensa2021}).
Our results provide insights into slow relaxation and metastability in many-body stochastic systems.

The remaining part of the paper is organized as follows.
In Sec.~\ref{sec:general}, we give a general formulation within a framework of classical Markov processes on discrete states.
We introduce eigenmode relaxation times in Sec.~\ref{sec:eigenmode}.
In Sec.~\ref{sec:metastability}, we numerically show that eigenmode expansion coefficients become huge for concrete models.
In Sec.~\ref{sec:non-interacting}, we consider non-interacting particles in a double-well potential.
Metastability in this model is due to large barrier of a single-particle potential, and hence we call it ``potential-induced metastability''.
In Sec.~\ref{sec:interacting}, we consider the same model with inter-particle interactions.
Metastability is induced by strong inter-particle interactions, and hence we call it ``interaction-induced metastability''.
In Sec.~\ref{sec:FA}, the FA model is investigated.
Metastability in this model is due to kinetic constraints and not of energetic origin.
It turns out that potential-induced metastability can be explained by the emergence of vanishingly small eigenvalues, whereas the other two are not associated with small eigenvalues: they are explained by many-body explosion of expansion coefficients.
Finally, we discuss our results in Sec.~\ref{sec:discussion}.

\section{General formulation}
\label{sec:general}
\subsection{Setup}
Let us consider a system with discrete states $n=1,2,\dots,D$.
The energy of the state $n$ is denoted by $E_n$.
The probability distribution $P_n(t)$ evolves following the master equation
\begin{equation}
\frac{dP_n(t)}{dt}=-\sum_{m=1}^DG_{n,m}P_m(t),
\label{eq:master}
\end{equation}
where the matrix $G$ is the generator of the stochastic dynamics.
For $n\neq m$, $G_{n,m}$ represents the transition rate from the state $m$ to $n$, and $G_{n,n}=-\sum_{m(\neq n)}G_{m,n}$.
When the system is coupled to a thermal bath at the inverse temperature $\beta$, $G$ satisfies the detailed-balance condition, $G_{n,m}e^{-\beta E_m}=G_{m,n}e^{-\beta E_n}$.

The detailed-balance condition ensures that $G$ is diagonalizable with real (and non-negative) eigenvalues since it is made symmetric via the similarity transformation $G\to G'=e^{\beta H/2}Ge^{-\beta H/2}$ with $H=\mathrm{diag}(E_1,E_2,\dots,E_D)$.
We here assume that the stationary state is unique and eigenvalues are ordered as $0=\Lambda_0<\Lambda_1\leq\Lambda_2\dots\leq\Lambda_{D-1}$.
The right and left eigenvectors for an eigenvalue $\Lambda_\alpha$ are respectively denoted by $\vec{\Phi}^{(\alpha)}$ and $\vec{\Pi}^{(\alpha)}$: $G\vec{\Phi}^{(\alpha)}=\Lambda_\alpha\vec{\Phi}^{(\alpha)}$ and $G^\mathrm{T}\vec{\Pi}^{(\alpha)}=\Lambda_\alpha\vec{\Pi}^{(\alpha)}$.

Let us normalize right eigenvectors by using the $L^1$ norm $\|\cdot\|_1$ as follows:
\begin{equation}
\|\vec{\Phi}^{(\alpha)}\|_1:=\sum_{n=1}^D|\Phi^{(\alpha)}_n|=1,
\label{eq:norm_right}
\end{equation}
where $\Phi^{(\alpha)}_n$ are $n$th component of $\vec{\Phi}^{(\alpha)}$.
We will later see that this normalization is convenient for our purpose.
Since the dual norm of the $L^1$ norm is the $L^\infty$ norm $\|\cdot\|_\infty$ in the sense that $|\vec{A}\cdot\vec{B}|\leq\|\vec{A}\|_\infty\|\vec{B}\|_1$ for any set of vectors $\vec{A}$ and $\vec{B}$, we normalize left eigenvectors by using $L^\infty$ norm as follows:
\begin{equation}
 \|\vec{\Pi}^{(\alpha)}\|_\infty:=\max_n|\Pi^{(\alpha)}_n|=1,
 \label{eq:norm_left}
 \end{equation}
 where $\Pi^{(\alpha)}_n$ are $n$th component of $\vec{\Pi}^{(\alpha)}$.

The mode $\alpha=0$ corresponds to the stationary state, and $\Phi^{(0)}_n=e^{-\beta E_n}/Z$ with $Z=\sum_ne^{-\beta E_n}$ and $\Pi^{(0)}_n=1$ for all $n$.
Without loss of generality, we choose the signs of $\vec{\Phi}^{(\alpha)}$ and $\vec{\Pi}^{(\alpha)}$ so that $\max_n\Pi^{(\alpha)}_n=1$ and $\vec{\Pi}^{(\alpha)}\cdot\vec{\Phi}^{(\alpha)}>0$.

The probability distribution $\vec{P}(t)=(P_1(t),\dots,P_D(t))^\mathrm{T}$ can be expanded in terms of right eigenvectors of $G$ as $\vec{P}(t)=\sum_\alpha C_\alpha e^{-\Lambda_\alpha t}\vec{\Phi}^{(\alpha)}$, where
\begin{equation}
C_\alpha=\frac{\vec{\Pi}^{(\alpha)}\cdot\vec{P}(0)}{\vec{\Pi}^{(\alpha)}\cdot\vec{\Phi}^{(\alpha)}}
\label{eq:coeff}
\end{equation}
is an eigenmode expansion coefficient in the initial state.

\subsection{Eigenmode relaxation time}
\label{sec:eigenmode}
Obviously, $\Lambda_\alpha$ gives the decay rate of $\alpha$th eigenmode, and hence the slowest decay rate is given by the spectral gap $\Lambda_1$ of $G$.
This observation indicates that slow relaxation should accompany small $\Lambda_1$.
Following this idea, in previous works~\cite{Gaveau1987, Gaveau1998, Biroli2001, Macieszczak2016}, metastability is characterized by the presence of small eigenvalues.
This argument implicitly assumes that $C_\alpha$ is not too large.
As we point out below, however, this assumption is typically not satisfied in many-body systems.

For a given initial state $\vec{P}(0)=\sum_\alpha C_\alpha\vec{\Phi}^{(\alpha)}$, let us define the relaxation time $\tau_\alpha$ for each eigenmode.
The expectation value of a physical quantity $\vec{O}=(O_1,O_2,\dots,O_D)^\mathrm{T}$ with $\|\vec{O}\|_\infty=1$ is given by
\begin{equation}
\braket{O}=\vec{O}\cdot\vec{P}(t)=\sum_\alpha C_\alpha e^{-\Lambda_\alpha t}\vec{O}\cdot\vec{\Phi}^{(\alpha)}.
\end{equation}
The contribution from $\alpha$th eigenmode is thus bounded as
\begin{equation}
|C_\alpha e^{-\Lambda_\alpha t}\vec{O}\cdot\vec{\Phi}^{(\alpha)}|\leq |C_\alpha|e^{-\Lambda_\alpha t}\|\vec{O}\|_\infty\|\vec{\Phi}^{(\alpha)}\|_1=|C_\alpha|e^{-\Lambda_\alpha t},
\end{equation}
where the normalization~(\ref{eq:norm_right}) is used.
Thus the contribution from $\alpha$th eigenmode is negligible when $|C_\alpha|e^{-\Lambda_\alpha t}\ll 1$, and hence it is appropriate to define the eigenmode relaxation time $\tau_\alpha$ as the time at which $|C_\alpha| e^{-\Lambda_\alpha t}=\delta$ for a fixed constant $\delta$.
Throughout the paper, we fix $\delta=0.5$.

We then have
\begin{equation}
\tau_\alpha=\frac{\ln\left(|C_\alpha|/\delta\right)}{\Lambda_\alpha},
\label{eq:tau}
\end{equation}
which implies that large expansion coefficients can cause the delay of the relaxation~\footnote{If we choose another normalization condition for right eigenvectors, the criterion of the relaxation $|C_\alpha|e^{-\Lambda_\alpha t}=\delta$ should be modified. For example, if we employ the normalization $\|\vec{\Phi}^{(\alpha)}\|_2:=\left(\sum_{n=1}^D|\Phi^{(\alpha)}_n|^2\right)^{1/2}=1$, we have $|\vec{O}\cdot\vec{\Phi}^{(\alpha)}|\sim D^{1/2}$, and hence the criterion of the relaxation should be $|C_\alpha|e^{-\Lambda_\alpha t}=\delta D^{-1/2}$ with $\delta$ being a fixed constant. Accordingly, the formula~(\ref{eq:tau}) should also be modified. In other words, Eq.~(\ref{eq:tau}) is appropriate only for the normalization scheme~(\ref{eq:norm_right})}.

Which eigenmodes are responsible for metastable states?
To answer this question, let us examine the possible largest value of $\tau_\alpha$ over all initial distributions.
From Eq.~(\ref{eq:coeff}), we have an upper bound 
\begin{equation}
|C_\alpha|\leq\frac{\|\vec{\Pi}^{(\alpha)}\|_\infty\|\vec{P}(0)\|_1}{\vec{\Pi}^{(\alpha)}\cdot\vec{\Phi}^{(\alpha)}}
=\frac{1}{\vec{\Pi}^{(\alpha)}\cdot\vec{\Phi}^{(\alpha)}}=:\Psi_\alpha.
\label{eq:Psi}
\end{equation}
This upper bound is always realizable: $|C_\alpha|=\Psi_\alpha$ when $P_n(0)$ is nonzero only for $n$ such that $\Pi^{(\alpha)}_n=1$.
Thus, we obtain
\begin{equation}
\max_{\vec{P}(0)}\tau_\alpha=\frac{\ln(\Psi_\alpha/\delta)}{\Lambda_\alpha}=:\tilde{\tau}_\alpha.
\label{eq:tau_tilde}
\end{equation}
The eigenmode relaxation times $\{\tilde{\tau}_\alpha\}$ tell us about which eigenmodes can compose metastable states.
In bra-ket notation, the time evolution operator is written as
\begin{equation}
e^{-Gt}=\sum_\alpha\Psi_\alpha e^{-\Lambda_\alpha t}\ket{\Phi^{(\alpha)}}\bra{\Pi^{(\alpha)}}.
\end{equation}
For a timescale specified by $t$, we consider a subset $\mathcal{M}$ of eigenmodes such that $\tilde{\tau}_\alpha\ll t$ for any $\alpha\notin\mathcal{M}$.
Since $\Psi_\alpha e^{-\Lambda_\alpha t}=\delta e^{\Lambda_\alpha(\tilde{\tau}_\alpha-t)}\ll 1$ for $\alpha\notin\mathcal{M}$, we have $e^{-Gt}\approx\sum_{\alpha\in\mathcal{M}}\Psi_\alpha e^{-\Lambda_\alpha t}\ket{\Phi^{(\alpha)}}\bra{\Pi^{(\alpha)}}$, which implies that a metastable state consists of eigenmodes in $\mathcal{M}$.

\begin{figure}[t]
\centering
\includegraphics[width=0.5\linewidth]{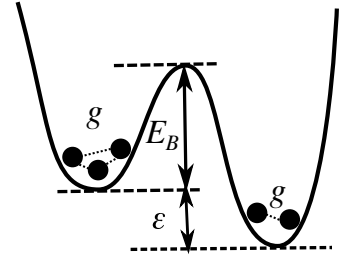}
\caption{Schematic picture of the model. Here, $E_B$ denotes the energy barrier and $\varepsilon$ the potential energy of the left well. In the interacting model, particles in the same potential well interact with each other ($g$ denotes the interaction strength).}
\label{fig:model}
\end{figure}

\section{Metastability and explosive growth of eigenmode expansion coefficients}
\label{sec:metastability}
In this section, we show that $\Psi_\alpha$ becomes huge and significantly changes the relaxation time in many-body stochastic models, i.e., an $N$-particle system in a double-well potential and the FA model.
Through the analysis of those models, we consider three kinds of metastability, i.e., the potential-induced metastability, the interaction-induced metastability, and the metastability due to kinetic constraints.
It is shown that the potential-induced metastability is associated with small eigenvalues, whereas the others are caused by an explosive growth of expansion coefficients $\Psi_\alpha$.

\subsection{Non-interacting particles in a double-well potential}
\label{sec:non-interacting}
Let us begin with the simplest model, i.e., $N$ independent particles in a double-well potential under the two-state approximation (see Fig.~\ref{fig:model}).
Each particle $i$ ($i=1,2,\dots,N$) is either in the left or right well, which is expressed by a spin variable $\sigma_i=\pm 1$ ($\sigma_i=1$ corresponds to the left well).
Then $N$-particle state is specified by a set of spin variables $(\sigma_1,\sigma_2,\dots,\sigma_N)$, and there are $D=2^N$ different states.
The state $\sigma_i=+1$ (left well) has the energy $\varepsilon>0$, and the energy barrier undergoing in the transition from $\sigma_i=+1$ to $-1$ is given by $E_B>0$.
By choosing the temperature as the unit of energy (i.e., $\beta=1$), transitions of $i$th particle from $\sigma_i=1$ to $-1$ and vice versa happen at rates given by the Arrhenius law $\tau_0^{-1}e^{-E_B}$ and $\tau_0^{-1}e^{-(E_B+\varepsilon)}$, respectively.
Here, $\tau_0$ is a certain microscopic timescale, which is chosen as the unit of time, i.e., $\tau_0=1$.
This transition rate satisfies the detailed-balance condition for the Hamiltonian $H=\varepsilon N_+:=\varepsilon\sum_{i=1}^N(\sigma_i+1)/2$, where the number of particles in the left well is denoted by $N_+$.
Similarly, we define $N_-=N-N_+$ as that in the right well.

In the single-particle problem, the generator is a two-by-two matrix, and it has two eigenvalues $\lambda_0=0$ and $\lambda_1=e^{-E_B}(1+e^{-\varepsilon})$.
The right and left eigenvectors are denoted by $\vec{\phi}^{(\alpha)}$ and $\vec{\pi}^{(\alpha)}$ ($\alpha=0$ or 1), respectively.
In the $N$-particle case, the eigenvalues $\{\Lambda_\alpha\}$ are given by the sum of single-particle eigenvalues as $\Lambda_\alpha=\sum_{i=1}^N\alpha_i\lambda_1$,
where $\alpha_i=0,1$ specifies which eigenmode is occupied by $i$th particle.
The corresponding eigenvectors are given by the product of single-particle eigenvectors:
$\vec{\Phi}^{(\alpha)}=\vec{\phi}^{(\alpha_1)}\otimes\vec{\phi}^{(\alpha_2)}\dots\otimes\vec{\phi}^{(\alpha_N)}$ and $\vec{\Pi}^{(\alpha)}=\vec{\pi}^{(\alpha_1)}\otimes\vec{\pi}^{(\alpha_2)}\dots\otimes\vec{\pi}^{(\alpha_N)}$.

The maximum expansion coefficient $\Psi_\alpha$ in Eq.~(\ref{eq:Psi}) for each $\alpha$ is explicitly calculated as follows:
\begin{equation}
\Psi_\alpha=\left(\frac{2}{1+e^{-\varepsilon}}\right)^{\sum_{i=1}^N\alpha_i}.
\end{equation}
When $\sum_{i=1}^N\alpha_i=O(N)$, $\Psi_\alpha=e^{O(N)}$ for $\varepsilon>0$.
Expansion coefficients diverge exponentially in $N$.
Although the corresponding eigenvalue is extensively large $\Lambda_\alpha=O(N)$, it is not correct that this eigenmode decays in a timescale $t\sim\Lambda_\alpha^{-1}=O(N^{-1})$.
The eigenmode relaxation time $\tilde{\tau}_\alpha$ is evaluated as
\begin{equation}
\tilde{\tau}_\alpha=-\frac{\ln(\delta)}{\Lambda_\alpha}+\frac{e^{E_B}}{1+e^{-\varepsilon}}\ln\left(\frac{2}{1+e^{-\varepsilon}}\right),
\end{equation}
whose second term is independent of the system size and $\Lambda_\alpha$.

In this model, the ``all-left'' state, i.e., $\sigma_1=\sigma_2=\dots=\sigma_N=1$, is considered to be a many-body metastable state for large $E_B$.
Expansion coefficients for this state are explicitly calculated, and it is found that $C_\alpha=\Psi_\alpha$ and $\tau_\alpha=\tilde{\tau}_{\alpha}$ for all $\alpha$, i.e., this state gives the maximum expansion coefficients for all eigenmodes.

Even in this simplest case, explosive growth of expansion coefficients occurs.
However, $\Psi_\alpha$ does not depend on $E_B$, and hence large $\Psi_\alpha$ is not related to metastability in this non-interacting model. 

\begin{figure}[t]
\centering
\includegraphics[width=0.9\linewidth]{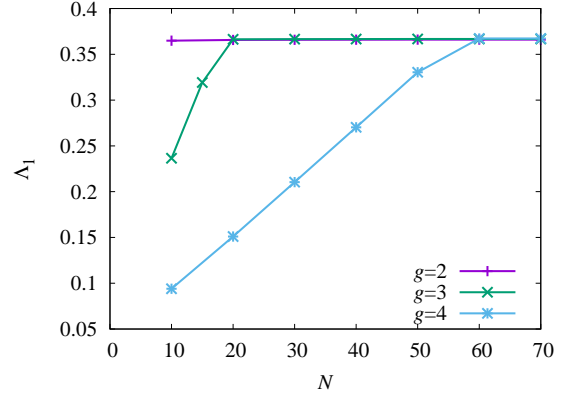}
\caption{Spectrum gap $\Lambda_1$ of $G$ against the system size $N$ for $g=2$, 3, and 4.
The saturated value of the gap at large $N$ is almost independent of $N$.}
\label{fig:gap}
\end{figure}

\begin{figure}[t]
\centering
\includegraphics[width=0.9\linewidth]{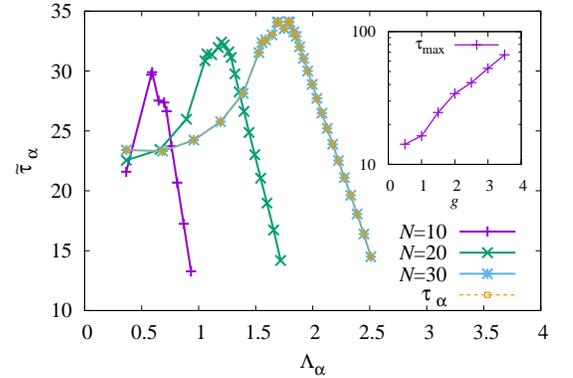}
\caption{Eigemmode relaxation times $\{\tau_\alpha\}$ against their eigenvalues $\Lambda_\alpha$ for different $N$. We find a peak structure, and the peak shifts towards larger eigenvalues as $N$ increases.
We also plot $\{\tilde{\tau}_{\alpha}\}$, which is indistinguishable from $\{\tau_\alpha\}$.
(Inset) Logarithmic plot of $\tau_\mathrm{max}$ against $g$ for $N=30$.}
\label{fig:tau}
\end{figure}

\begin{figure}[t]
\centering
\includegraphics[width=0.9\linewidth]{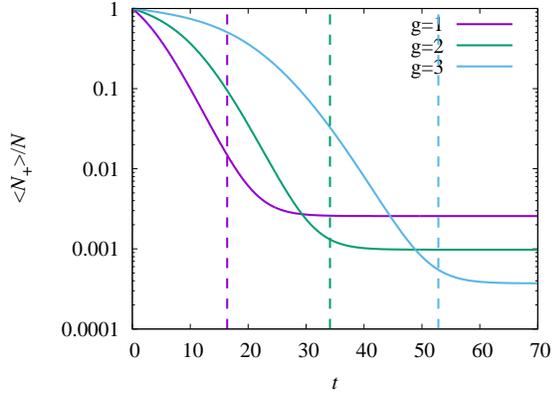}
\caption{Dynamics of the expectation value of $N_+$ for different values of $g$ at $N=30$.
Vertical dashed lines represent $\tau_\mathrm{max}$ for each $g$.}
\label{fig:dynamics}
\end{figure}

\subsection{Interacting particles in a double-well potential}
\label{sec:interacting}
Let us introduce interactions.
We here consider a simple model in which two particles in the same potential well equally interact with each other.
The interaction energy is given by $E_\mathrm{int}=-(g/N)[N_+(N_+-1)/2+N_-(N_--1)/2]$, where $g>0$ denotes the interaction strength (the scaling $1/N$ ensures the extensivity of the energy).
The interaction changes the effective potential felt by a single particle: the energy is written as $E(N_+)=\varepsilon_+N_++\varepsilon_-N_-$ with $\varepsilon_+=\varepsilon-g(N_+-1)/N$ and $\varepsilon_-=-g(N_--1)/N$.
The effective energy barrier from $\sigma_i=\pm 1$ to $\mp 1$ is thus given by $E_B^{(\pm)}=E_B+\varepsilon-\varepsilon_{\pm}$.
The model with interactions is obtained by replacing $(\varepsilon,E_B)$ in the non-interacting model by $(\varepsilon_+-\varepsilon_-,E_B^{(+)})$.
Obviously, interactions $g>0$ effectively increases the energy barrier and causes slower relaxation, which we call interaction-induced metastability.

The master equation in this model is generated by a $2^N$-dimensional generator.
Instead of considering the full generator, let us employ the simplified description by focusing on the dynamics of a collective variable $N_+\in\{0,1,\dots,N\}$ only.
Since the energy only depends on $N_+$ in our model, we can exactly obtain the dynamics of $N_+$.
Let the probability distribution of $N_+$ at time $t$ be denoted by $P_{N_+}(t)$.
Then the master equation for $P_{N_+}(t)$ is written as
\begin{multline}
\frac{dP_{N_+}(t)}{dt}=-G_{N_+,N_++1}P_{N_++1}(t)-G_{N_+,N_+-1}P_{N_+-1}(t)\\
+(G_{N_++1,N_+}+G_{N_+-1,N_+})P_{N_+}(t),
\label{eq:master_N}
\end{multline}
where $G_{N_+\mp 1,N_+}=-N_{\pm}e^{-E_B^{(\pm)}}$.
This generator satisfies the detailed-balance condition of the form
\begin{equation}
G_{N_+-1,N_+}e^{-F(N_+)}=G_{N_+,N_+-1}e^{-F(N_+-1)},
\end{equation}
where $F(N_+)=E(N_+)-S(N_+)$ denotes the free energy of a state specified by $N_+$.
The entropy $S(N_+)$ is given by $S(N_+)=\ln(N!/(N_+!N_-!)$.

The $(N+1)$-dimensional generator in Eq.~(\ref{eq:master_N}) is a diagonal block of the original $2^N$-dimensional full generator.
Thus, the full generator has other $2^N-(N+1)$ eigenmodes, which are dropped in this description.
However, it does not affect the conclusion (see Appendix~\ref{sec:full}).

We now fix $E_B=1$ and $\varepsilon=5$.
In our model, there is the non-ergodic phase for $g\gtrsim 8.8$, in which we have an extensive free energy barrier $\Delta F\propto N$ and the relaxation time $\tau_\mathrm{rel}$ diverges in the thermodynamic limit as $\tau_\mathrm{rel}=e^{\Delta F}=e^{O(N)}$.
We discuss the non-ergodic phase in Appendix~\ref{sec:non-ergodic}, and here we focus on the ergodic phase at which the free energy $F(N_+)$ is a monotonic function of $N_+$.

First let us investigate the $g$-dependence of the spectral gap of $G$.
Numerical results for different $g$ and $N$ are presented in Fig.~\ref{fig:gap}.
It clearly shows that the spectral gap of $G$ is almost independent of the value of $g$ for large $N$, although the finite-size effect is stronger for larger $g$.
It means that the spectrum gap $\Lambda_1$ does not reflect the increase of the relaxation time due to interactions.

Next, we compute $\{\tilde{\tau}_\alpha\}$ at $g=2$ for different $N$ and show the result in Fig.~\ref{fig:tau}.
Although the typical value of $\Lambda_{\alpha}$ linearly grows with $N$, that of $\tilde{\tau}_\alpha$ looks convergent for large $N$.
According to Eq.~(\ref{eq:tau_tilde}), it means that expansion coefficients typically grows exponentially in $N$, which is already seen in the non-interacting model.
In Fig.~\ref{fig:tau}, $\tau_\alpha$ for the all-left initial state is also plotted by squares with a dashed line in Fig.~\ref{fig:tau}, which completely agrees with $\tilde{\tau}_\alpha$ as in non-interacting case.
The all-left state has the maximum expansion coefficients for all eigenmodes, $|C_\alpha|=\Psi_\alpha$.

It turns out that the eigenmode $\alpha^*$ that gives the largest relaxation time $\tau_\mathrm{max}=\max_\alpha\tilde{\tau}_\alpha=\tilde{\tau}_{\alpha^*}$ appears in the middle of the spectrum, $\Lambda_{\alpha^*}\propto N$, not at the spectrum edge.
Since the relaxation particularly slows down far from equilibrium with large $N_+$, an extensive number of particles simultaneously undergo dissipation in a many-body metastable state, which would be the reason why eigenmodes with extensively large eigenvalues are important here.

We plot $\tau_\mathrm{max}$ against $g$ in the inset of Fig.~\ref{fig:tau}.
We find that $\tau_\mathrm{max}$ increases exponentially in $g$, which is consistent with the intuitive picture that interactions effectively increase the energy barrier.
This increase of the relaxation time stems from rapid growth of expansion coefficients with $g$.

In order to compare the physical relaxation time with $\tau_\mathrm{max}$, we numerically compute the dynamics of $\braket{N_+}=\sum_{N_+}N_+P_{N_+}(t)$ starting from the all-left initial state.
The numerical result for different $g$ at $N=30$ is shown in Fig.~\ref{fig:dynamics}.
The quantity $\braket{N_+}$ reaches the stationary value roughly at $t=\tau_\mathrm{max}$, which is shown by vertical dashed lines in Fig.~\ref{fig:dynamics}.
It is thus confirmed that $\tau_\mathrm{max}$ coincides with the physical relaxation time of the system.

\begin{figure}[t]
\centering
\includegraphics[width=0.9\linewidth]{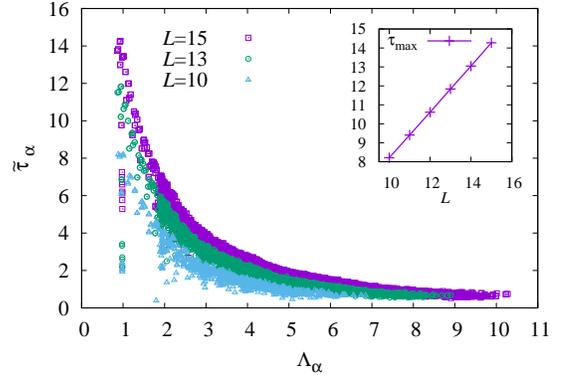}
\caption{Eigemmode relaxation times $\{\tilde{\tau}_\alpha\}$ against their eigenvalues $\Lambda_\alpha$ for different $L$ in the FA model with $\rho=0.1$.
(Inset) $\tau_\mathrm{max}$ against $L$.}
\label{fig:FA}
\end{figure}

\subsection{Fredrickson-Andersen model}
\label{sec:FA}
In the model considered so far, interaction-induced metastability is explained by the many-body explosion of expansion coefficients.
Here we examine the FA model~\cite{Fredrickson1984} as a prototypical example of kinetically constrained models~\cite{Ritort2003} studied in glass and jamming transitions~\cite{Berthier2011}.
Metastability in a kinetically constrained model is not due to energy barriers but due to dynamical rules.
As we see below, the many-body explosion of expansion coefficients also explains this kind of metastability.

Suppose that each site $i\in\{1,2,\dots,L\}$ is occupied ($n_i=1$) or empty ($n_i=0$), and the transition rates from $n_i=0$ to 1 and that from $n_i=1$ to 0 are given by $C_i\rho$ and $C_i(1-\rho)$, respectively, where $\rho$ corresponds to the average density at equilibrium and $C_i$ expresses dynamical constraints.
In the FA model~\cite{Fredrickson1984}, $C_i=1-n_{i-1}n_{i+1}$, which means that the site $i$ can change its state only if at least one of the two neighbors $i-1$ and $i+1$ is empty.
Since this transition rate satisfies the detailed balance condition with respect to the non-interacting Hamiltonian $H=-\mu\sum_{i=1}^Ln_i$, where $\mu=\ln[\rho/(1-\rho)]$ denotes the chemical potential, the equilibrium state is trivial but its dynamics exhibits interesting properties, which is a common characteristic of kinetically constrained models.

In the FA model, it is rigorously proved that the spectral gap of $G$ remains finite in the thermodynamic limit for any $0\leq\rho<1$~\cite{Cancrini2008}.
However, the maximum relaxation time exhibited by this model diverges in the thermodynamic limit for any $\rho$.
Because of the kinetic constraint, a cluster of particles can decay or grow only from its edges.
Therefore, if the initial state has a macroscopically large cluster of size $O(L)$, the relaxation time is at least of $O(L)$. 
A state with macroscopically large clusters is therefore regarded as a metastable state.
This increase of the maximum relaxation time should be due to an exponential growth of $\Psi_\alpha$~\footnote{At $\rho=1/2$, the Hamiltonian vanishes and $G$ is a symmetric matrix. Even in this case, the many-body explosion of expansion coefficients occurs.}.

Under the periodic boundary condition, $\{\tilde{\tau}_\alpha\}$ for $\rho=0.1$ are computed and plotted in Fig.~\ref{fig:FA}.
We see that in the FA model $\Psi_\alpha=e^{O(L)}$ for $\alpha$ with $\Lambda_\alpha=O(1)$, which results in $\tau_\mathrm{max}\propto L$ even though the spectral gap remains finite.

\section{Discussion}
\label{sec:discussion}
We have shown that the many-body explosion of eigenmode expansion coefficients is responsible for slow relaxation in many-body stochastic systems.
We emphasize that this is a generic phenomenon, and it would also occur in nonequilibrium dynamics without detailed balance.
Concepts related to slow relaxation, such as metastable states, must be reconsidered.

Since the maximum value $\Psi_\alpha$ of the expansion coefficient for an eigenmode $\alpha$ is given by the inverse of the overlap between the left eigenvector $\vec{\Pi}^{(\alpha)}$ and the right eigenvector $\vec{\Phi}^{(\alpha)}$, the many-body explosion of $\Psi_\alpha$ indicates a mismatch between the supports of these two vectors.
Under the detailed balance condition, left and right eigenvectors are given by $\vec{\Phi}^{(\alpha)}\propto e^{-\beta H/2}\vec{W}^{(\alpha)}$ and $\vec{\Pi}^{(\alpha)}\propto e^{\beta H/2}\vec{W}^{(\alpha)}$, where $\vec{W}^{(\alpha)}$ is the eigenvector of the symmetric matrix $G'=e^{\beta H/2}Ge^{-\beta H/2}$ with the eigenvalue $\Lambda_\alpha$.
Therefore, the support of $\vec{\Phi}_\alpha$ ($\vec{\Pi}^{(\alpha)}$) tends to be localized at low (high) energies, which may result in an extremely small overlap $\vec{\Pi}^{(\alpha)}\cdot\vec{\Phi}^{(\alpha)}=e^{-O(N)}$.
We however point out that explosive growth of $\Psi_\alpha$ occurs even when $G$ is a symmetric matrix, i.e., $H=0$, which is realized in the FA model with $\rho=1/2$.
We also emphasize that it would also occur in nonequilibrium stochastic dynamics without detailed balance.

In the two models examined in this work, $\{\tilde{\tau}_\alpha\}$ are broadly distributed, and no clear separation in $\{\tilde{\tau}_\alpha\}$ is observed.
In such a case, it would be a difficult problem to identify metastable states~\cite{Tanase-Nicola2004, Kurchan2011, Jack2013}.
Existing mathematical theory on spectral characterization of metastable states~\cite{Gaveau1998} provides a recipe to construct a metastable state as a linear combination of relevant eigenmodes of $G$, but it is not applicable to our case, in which slow relaxation is caused by the many-body explosion of expansion coefficients.
Although we can identify which eigenmodes are important for metastable states by looking at $\{\tilde{\tau}_\alpha\}$, it is not straightforward to find a systematic method to construct metastable states by using them.
We leave it as a future problem.

\begin{acknowledgments}
The author thanks Tatsuhiko Shirai for carefully reading the manuscript.
This work was supported by JSPS KAKENHI Grant Numbers JP19K14622, JP21H05185.
\end{acknowledgments}

\appendix
\section{Diagonalization of full generator}
\label{sec:full}
In the main text, we consider the model of $N$ interacting particles in a double-well potential within the two-state approximation.
In this model, each microscopic state is specified by the set of ``spin'' variables $\{\sigma_i\}=\{\sigma_1,\sigma_2,\dots,\sigma_N\}$, where $\sigma_i=+1$ ($-1$) means that $i$th particle is in the left (right) potential well.
In the main text, the master equation for the probability distribution $P_{N_+}(t)$ of the collective variable $N_+=\sum_{i=1}^N(\sigma_i+1)/2$ is examined.

\begin{figure}[h]
\centering
\includegraphics[width=\linewidth]{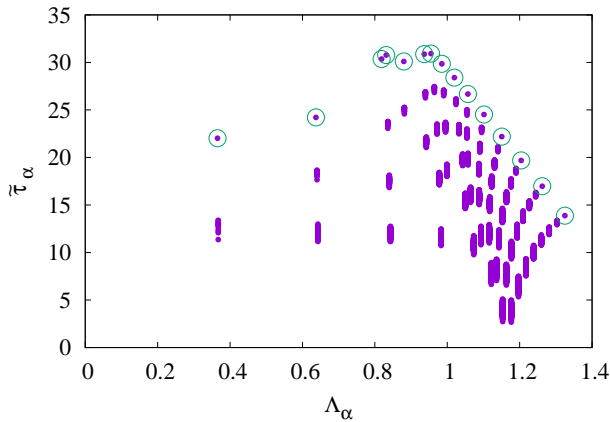}
\caption{Eigenmode relaxation times $\{\tilde{\tau}_\alpha\}$ obtained by the diagonalization of the full generator of the master equation for $N=15$, $g=2$, $E_B=1$, and $\varepsilon=5$.
Those obtained by using the generator of the master equation for $P_{N_+}(t)$ are shown by open circles.}
\label{fig:full}
\end{figure}

Here we diagonalize the generator of the master equation for the probability distribution $P_{\{\sigma_i\}}(t)$ over all microscopic states, which we call ``the full generator'', and compare with the description using the collective variable $N_+$.
The result for $N=15$, $g=2$, $E_B=1$, and $\varepsilon=5$ is shown in Fig.~\ref{fig:full}.
In the calculation, in order to avoid eigenvalue degeneracies, we introduced very weak inhomogeneity: $\varepsilon$ is replaced by $\varepsilon_i=\varepsilon+\delta_i$, where $\{\delta_i\}$ are independent random variables uniformly drawn from $[0,10^{-4}]$.

Figure~\ref{fig:full} shows that the full generator contains all the eigenmodes of the generator of the master equation for the collective variable (shown by open circles).
This is because the collective variable is completely decoupled from the other degrees of freedom in our model.
In general, of course, coarse-grained description using collective variables is not exact.

\section{Non-ergodic phase}
\label{sec:non-ergodic}
In the two-state model of interacting particles in the double-well potential, we have the non-ergodic phase for $g>g_c$, where $g_c\approx 8.8$ for $\varepsilon=5$.
When $g<g_c$, $F(N_+)$ is a monotonic function of $N_+$, whereas when $g>g_c$, it becomes non-monotonic and has a local minimum, which is due to the mean-field character of the model.
The non-ergodicity in the thermodynamic limit is a result of an extensive free energy barrier $\Delta F\propto N$.
The local minimum is interpreted as a metastable state in this case, and its lifetime grows with $N$ as $e^{\Delta F}=e^{O(N)}$.

\begin{figure}[t]
\centering
\includegraphics[width=\linewidth]{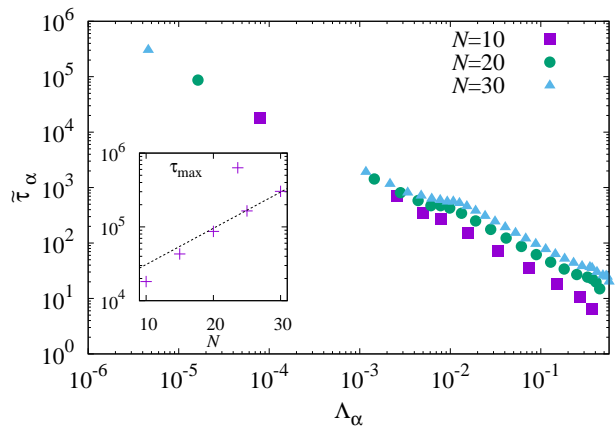}
\caption{Eigenmode relaxation times $\{\tilde{\tau}_\alpha\}$ against $\{\Lambda_\alpha\}$ in the non-erogdic phase. (Inset) $\tau_\mathrm{max}=\max_\alpha\tilde{\tau}_\alpha$ against $N$. It increases exponentially in $N$, in agreement with $\text{const.}\times e^{\Delta F}$ that is shown by the dashed line.
The parameters are chosen as $\varepsilon=5$, $E_B=1$, and $g=10$.}
\label{fig:nonergodic}
\end{figure}

In the non-ergodic phase, metastability is fully explained by the gap closing of $G$.
We find that $\alpha^*=1$ and $\Lambda_1\propto e^{-\Delta F}$.
The corresponding maximum expansion coefficient $\Psi_1$ does not grow with $N$.
This is explained by the fact that relaxation of a metastable state across the free energy barrier is regarded as a single-particle problem with the cooridinate $x=N_+/N$.
There is no ``many-body'' explosion in the relevant expansion coefficient.
We show numerical results for $\varepsilon=5$, $E_B=1$, and $g=10>g_c$ in Fig.~\ref{fig:nonergodic}.
The maximum eigenmode relaxation time $\tau_\mathrm{max}=\max_\alpha\tilde{\tau}_\alpha$ agrees with $\text{const.}\times e^{\Delta F}$, where $\Delta F\approx 0.113$ (see the inset of Fig.~\ref{fig:nonergodic}.

\bibliography{apsrevcontrol,physics}

\end{document}